\documentclass[a4paper,11pt]{article}

\usepackage[top=12mm,bottom=12mm,left=30mm,right=30mm,head=12mm,includeheadfoot]{geometry}
\usepackage{subcaption}
\usepackage[utf8]{inputenc}
\usepackage[english]{babel}
\usepackage[T1]{fontenc}
\usepackage{graphicx}
\usepackage{comment}
\usepackage[separate-uncertainty=true]{siunitx}
\DeclareSIUnit\bar{bar}
\usepackage{amsmath, amssymb}
\usepackage{upgreek}
\usepackage{soul}
\usepackage{xcolor}
\usepackage{authblk}
\usepackage[version=4]{mhchem}
\usepackage{float}
\usepackage[]{hyperref}

\title{Temperature-insensitive tunable and stable Fabry-Perot cavity for atomic physics}
\author[1,*]{Joshua Ruelle}
\author[1]{Martin Hauden}
\author[1]{Francisco S. Ponciano-Ojeda}
\author[1]{Marion Delehaye}
\affil[1]{Université Marie et Louis Pasteur, SUPMICROTECH, CNRS, Institut FEMTO-ST, F- 25000 Besançon, France}
\affil[*]{corresponding author: joshua.ruelle@femto-st.fr}
\date{}

\begin{document}
\maketitle

\begin{abstract}
	Optical Fabry-Perot cavities are crucial tools for metrology experiments, where they achieve extreme length stability and now can serve as highly proficient references for probing fine atomic transitions. For some atomic physics experiments however, tunability of the system to atomic transitions enables atom-light interactions. Achieving both frequency stability and tunability in a single cavity has remained a challenge, forcing metrology experiments exploiting atom-cavity interactions to rely on external active feedback systems to stabilize the length of the cavity. Here, we describe a piezoelectrically-tunable cavity with a cancellation of the coefficient of thermal expansion at around $5^\circ\mathrm{C}$, achieving a fractional frequency instability of $4\times 10^{-13}$ level for 1~s integration time. This advance eliminates the need for external stabilization in many atom-cavity experiments, making this design ideal for applications such as ultra-stable superradiant lasers and other cavity quantum electrodynamics experiments.
\end{abstract}

\section{Introduction}
\label{sec:introduction}

Optical Fabry-Perot (FP) cavities are foundational to many cutting-edge systems in modern physics. Monolithic, ultra-stable FP cavities, achieving fractional frequency instabilities as low as $\sigma_y = 2.5 \times 10^{-17}$~\cite{lee_frequency_2026, Matei2017}, are critical components of optical clocks, where they provide short-term frequency stability. Meanwhile, piezoelectrically-tunable cavities hosting cold atoms are used to mediate light-matter interactions in cavity quantum electrodynamics (cQED) systems.
Recently, FP cavities have found a novel and challenging application at the intersection of time and frequency metrology and cQED: as the backbone of superradiant (SR) lasers operating as active optical clocks~\cite{Meiser2009, Norcia2017}, with the potential of achieving frequency instabilities as low as $4\times 10^{-18} (\tau/\mathrm{s})^{-1/2}$~\cite{kazakov_ultimate_2022}. In these systems, an ensemble of atoms is coupled to an optical cavity, which must be precisely tuned to the resonant frequency of a narrow atomic transition.
The atomic ensemble collectively emits a narrow-linewidth laser signal, with the emitted frequency remaining relatively insensitive to cavity detunings, thereby reducing the impact of cavity-induced frequency fluctuations by typically a factor $10^5$~\cite{Kuppens1994, Norcia2016}. Consequently, an ideally-performing superradiant laser targeting a fractional frequency instability of $4\times 10^{-18}$ would only require its optical cavity to achieve a fractional frequency instability on the order of $10^{-13}$.
Nevertheless, the residual fractional frequency instability of the cavity must still meet demanding standards, exceeding those typically required for conventional cQED experiments. The cavity must also be tunable to maintain resonance with the atomic transition, which introduces further technical complexity. Achieving simultaneously frequency stability and tunability, two properties that are typically difficult to reconcile in a single device, thus represents a unique challenge.

Beyond its application in SR lasers, tunable and stable cavities are also highly relevant for other precision experiments, such as cavity-enhanced spectroscopy~\cite{Axner_2014, saavedra_tunable_2021}, or gravitational-wave detection: large, stable and tunable FP cavities enable applications such as frequency-dependent squeezing in the LIGO-VIRGO collaboration~\cite{Kimble_2001, Ganapathy_2023, Acernese_2023}, or  space-borne detectors like the eLISA mission~\cite{Danzmann_2003}, for which tunable and stable cavities have already been investigated~\cite{mohle_highly_2013}.

In this context, we present the design of a piezoelectrically-tunable and stable FP cavity developed for an ytterbium (\ce{Yb}) SR laser experiment. This paper is organized as follows: we first describe the design of the cavity and its materials, then the metrological methods that we use to characterize its performance, and finally the achieved results, including optical, thermal, electrical properties as well as its stability that reaches the mid-$10^{-13}$ range.

\section{Composite cavity design}
\label{sec:design}

\noindent
\begin{minipage}[t]{0.49\linewidth}
	\centering
	\includegraphics[width=\linewidth]{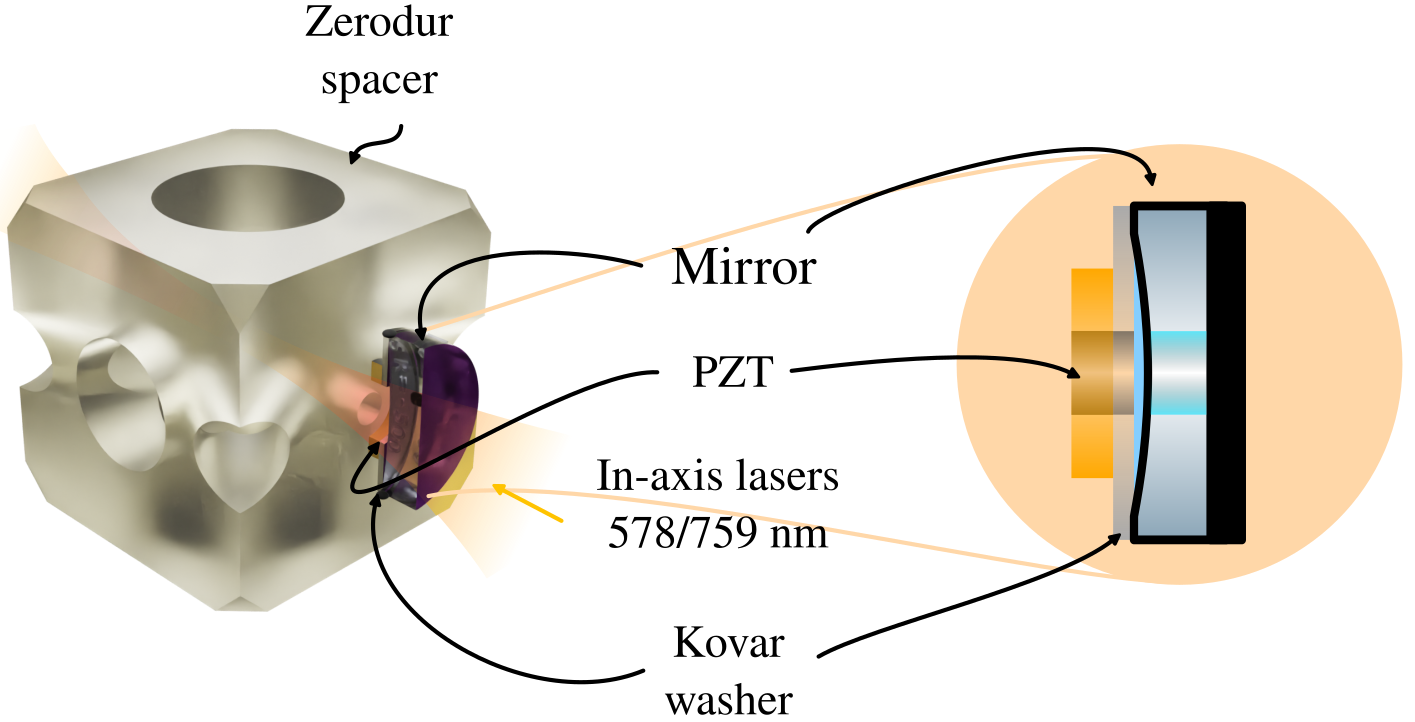}
	\captionsetup{width=\linewidth}
	\captionof{figure}{3D representation of the tunable Fabry-Perot cavity consisting of a 50-mm long Zerodur spacer and mirrors stacked on a PZT ring and a Kovar washer. }
	\label{fig:cavity_render}
\end{minipage}%
\hfill
\begin{minipage}[t]{0.49\linewidth}
	\centering
	\includegraphics[width=\linewidth]{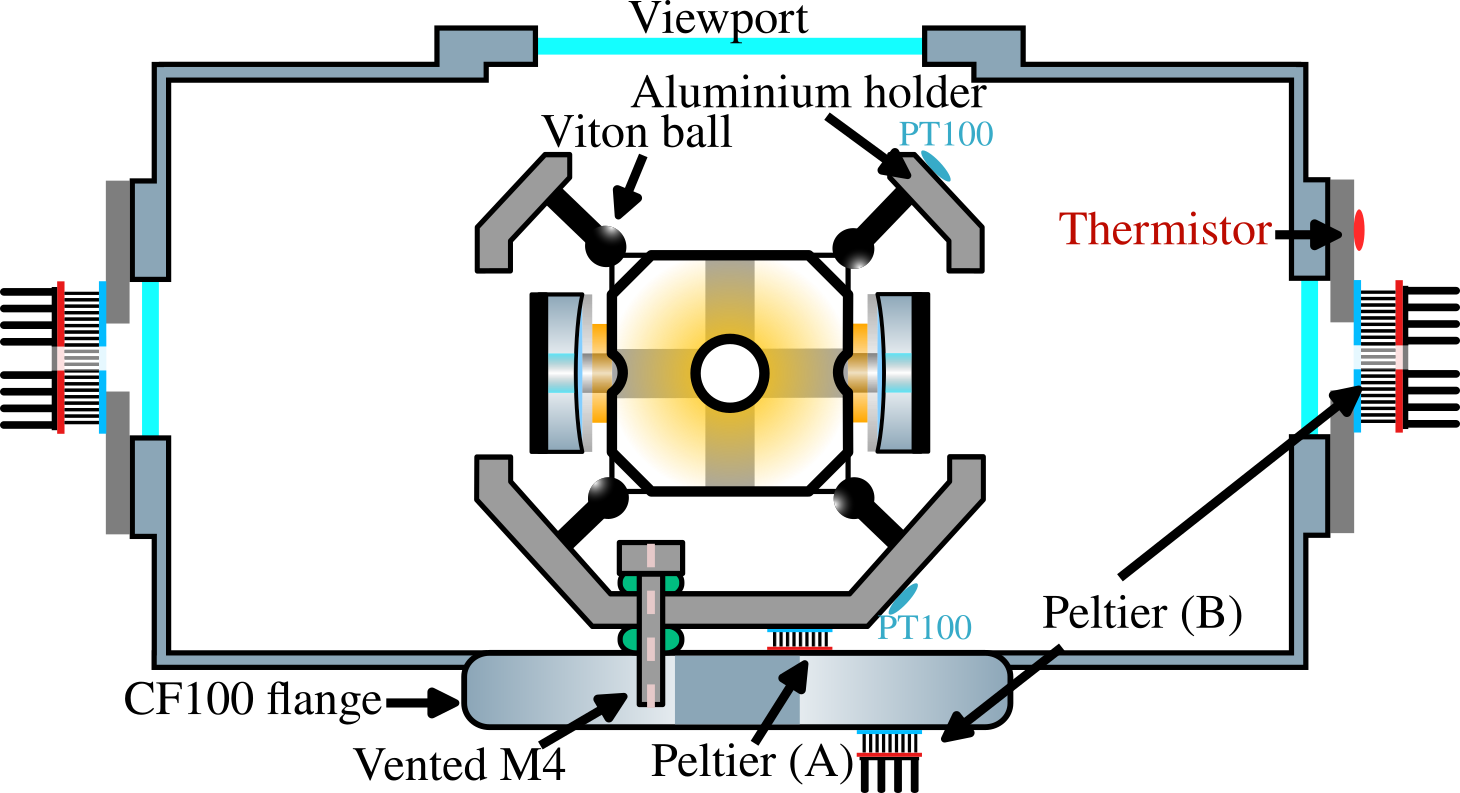}
	\captionsetup{width=\linewidth}
	\captionof{figure}{Schematic side-view of the cavity vacuum chamber.}
	\label{fig:cavity_thermics_drawing}
\end{minipage}

\vspace{0.2cm}

For a superradiant laser, the cavity must satisfy four key criteria: \textbf{(i)} \emph{optical access} , to enable the trap loading and imaging of atoms; \textbf{(ii)} \emph{tunability}, to ensure resonance with the atomic transition; \textbf{(iii)} \emph{frequency stability} , not to degrade the SR laser stability; and \textbf{(iv)} \emph{high finesse}  for suitable atom-cavity coupling.

To address these challenges, we developed  a cavity design including a $L_\mathrm{ZD}= \qty{50}{mm}$-long cubic cavity spacer in Zerodur inspired by the design of the National Physical Institute Laboratory (NPL) in London~\cite{Webster2011}. We mounted the spacer using a tetrahedral configuration, where four Viton balls are positioned at non-adjacent, diagonally opposed vertices of the cubic spacer. We modified the spacer to increase both the size and number of apertures \textbf{(i)}: the venting holes were enlarged to 20 and 28~mm, and 13-mm diameter holes have been added along the planar diagonal to allow for future improved atomic trapping (see Fig.~\ref{fig:cavity_render}).
Using finite-element methods similar to those in~\cite{Webster2011}, we optimized the depth of the corner cuts to 7.4~mm to minimize sensitivity to fluctuations of the holding forces. These simulations also estimate the acceleration sensitivity to less than $ 10^{-10}\,\mathrm{/(m/s^2)}$ along all three axes.

Tunability \textbf{(ii)} is ensured using a pair of $L_\mathrm{PZT} =2$~mm-thick lead zirconate titanate (PZT) transducers. With a free stroke of $3.3~\upmu$m for a maximum voltage of 200~V, these PZTs allow the cavity to cover several free-spectral ranges (FSR).

Regarding frequency stability \textbf{(iii)}, the PZT actuators can degrade the stability by two different mechanisms. First, they convert electrical noise into mechanical length noise that directly translates into frequency instability. Thorough electrical filtering is therefore essential. Second, they exhibit thermal expansion, characterized by a coefficient of thermal expansion $\alpha = \frac{1}{\Delta T}\frac{\Delta L}{L}$, where $\Delta L/L$ is the fractional length change induced by a temperature variation $\Delta T$. The coefficient of thermal expansion of the 2 mm-thick PZT actuators can be estimated  between \qty{0}{\celsius} and \qty{50}{\celsius}~\cite{mangeot2016full} at around $\alpha_\mathrm{PZT}(T)=a_\mathrm{PZT}(T-T_\mathrm{PZT})$, with $a_\mathrm{PZT} = (\num{-1.0\pm0.2})\times 10^{-7}~\mathrm{K^{-2}}$ and $T_\mathrm{PZT} = \qty{-0.9}{\celsius}$, resulting in  $\alpha_\mathrm{PZT}(T=\qty{25}{\celsius}) = -2.6 \times 10^{-6}~\mathrm{K^{-1}}$. Based on this estimation, we add between the PZT and the mirrors a pair of $L_\mathrm{Kov} = 1$~mm-thick Kovar washers, which have a coefficient of thermal expansion of $\alpha_\mathrm{Kov} = 5.86\times 10^{-6}~\mathrm{K^{-1}}$ between \qty{25}{\celsius} and \qty{100}{\celsius}.  
The Zerodur spacer has a negligible coefficient of thermal expansion $\alpha_\mathrm{ZD} = 0\qty{\pm5e-8}{\kelvin^{-1}}$ between \qty{0}{\celsius} and \qty{50}{\celsius}.

With these materials, the overall thermal noise floor is estimated at approximately $3\times 10^{-15}$~\cite{numata_thermal-noise_2004, Kessler2012}, with roughly half of this contribution originating from the PZTs.

The cavity is designed to support a SR laser based on ytterbium atoms \textbf{(iv)}. Consequently, the mirrors are highly-reflective at both the clock transition  wavelength ($ \ce{^1S_0} \leftrightarrow \ce{^3P_0}$) at \qty{578}{nm}, and the so-called \emph{magic wavelength}  at \qty{759}{nm}, where the light shifts of the  $\ce{^1S_0}$ and $\ce{^3P_0}$ states are equal. The mirrors are identical, plano-concave with a \qty{50}{cm} radius of curvature. The cavity is maintained under ultra-high vacuum conditions ($\leq \qty{e-11}{\milli\bar}$).

\section{Methods}
\label{sec:methods}
To assess the performance of the cavity, we measure the frequency of a \qty{578}{nm} laser\footnote{Toptica DL-SHG Pro.}, which is frequency-locked to the FP cavity using the Pound-Drever-Hall (PDH) technique~\cite{Drever_1983, black_introduction_2001}. A \qty{25}{MHz} EOM\footnote{Qubig PM7-VIS\_25.} is used to generate sidebands, and we detect the reflected signal, demodulate it, and feed it to an ultra-fast PID\footnote{Toptica FALC Pro.} to stabilize the laser frequency to the cavity resonance, achieving a lock bandwidth of roughly \qty{600}{\kilo\hertz}. The stabilized laser is then compared to an active hydrogen maser (H-maser) using an optical frequency comb\footnote{Toptica DFC.} to transfer fractional frequency stability from radio frequencies (RF) to the optical frequency domain. The OFC is locked on the H-maser with a fractional frequency instability $\sigma_y (\tau)= 10^{-13}(\tau/\mathrm{s})^{-1/2}$ for all integration times $\tau$ between \num{1} and \qty{e5}{s}. The beatnote between the FP cavity-stabilized laser and the optical frequency comb is detected using a photodiode and mixed with an RF signal generated by a referenced Digital Direct Synthesizer (DDS) to produce a beatnote between 9.5 and 10.5 MHz for efficient filtering. We finally acquire and analyze the beatnote with a frequency counter\footnote{K+K Messtechnik.}.

\section{Performance}
\label{sec:performance}

\subsection{Finesse measurement}
To characterize the optical performance of the cavity, we employed the ring-down technique~\cite{Rempe_1992} to measure its finesse at \qty{578}{nm}. Using a \qty{105}{ns} fall-time photodiode, we initially measured a decay time of $\qty{1.2\pm0.4}{\micro\second}$ prior to the bake-out of the vacuum chamber, corresponding to a finesse of $\num{21000\pm7000}$. However, the mirror quality degraded during the bake-out process, resulting in a final decay time of $\qty{411\pm 2}{ns}$ and a reduced finesse of $\num{6920 \pm 40}$. Our hypothesis is a possible contamination from the Non-Evaporable Getter (NEG) pump that was activated during the bake-out. This could be avoided by careful epoxy placement on the inside of the elements, careful activation of the ion-getter pump to avoid abrupt current changes, slower temperature ramp during bake-out and more thorough cleaning of the mirror substrates.

\subsection{Noise induced by PZT actuators}

The cavity tunability is achieved using ring PZT actuators\footnote{Noliac NAC2124.} with a thickness of 2~mm and inner/outer diameters of $\num{9}$ and \qty{15}{mm}, respectively. We measured a piezoelectric cavity frequency sensitivity of \qty{233}{\mega\hertz\per\volt}. This high sensitivity allows one FSR coverage with an applied voltage of only \qty{12}{V} but this induces also an increase of the absolute voltage noise influence on the cavity frequency.  

The PZT actuators show mechanical resonances at approximately \qty{110}{kHz} and $10~\text{MHz}$, attributed to global mechanical transverse modes (due to misalignment of ceramic grains) and bulk transverse/longitudinal modes, respectively. The voltage noise of our most stable power supply\footnote{Aim TTI EL155R.} was measured at $\qty{1}{\micro\volt\per\sqrt{\hertz}}$ at \qty{1}{Hz} for \qty{1}{V} output, which would result in a frequency noise of $\qty{0.2}{\kilo\hertz\per\sqrt{\hertz}}$ at \qty{1}{\hertz}. This noise level limits the fractional frequency stability of the cavity to approximately $\num{5e-13}$~\cite{numata_thermal-noise_2004}. Additionally, the residual voltage noise of $\qty{30}{\nano\volt\per\sqrt{\hertz}}$ at \qty{110}{\kilo\hertz} excited the mechanical resonances of the PZT, which were visible in the beatnote spectrum. We therefore designed a DC low-pass filter ($ R = \qty{100}{\mega\ohm} $, $ C = \qty{10}{\micro\farad} $) with a cutoff frequency at \qty{160}{\micro\hertz}, similar to the design in~\cite{mohle_highly_2013}.
For applications requiring fast actuation, the filter is AC-bypassed  using a secondary filter ($ R = \qty{1.2}{\kilo\ohm} $, $ C = \qty{88}{\nano\farad} $) providing a \qty{1.6}{\kilo\hertz} cutoff frequency for this branch. With the DC filter, the voltage noise has been reduced to \qty{200}{\nano\volt\per\sqrt{\hertz}}, corresponding to a flicker contribution of $\num{1e-13}$ to the cavity fractional frequency instability.

\subsection{Temperature control}

The temperature stability of the cavity is achieved through a combination of active temperature regulation and passive thermal low-pass filtering.
More specifically, the cavity is tetrahedrically-mounted via Viton balls within an aluminium holder under vacuum, which is  attached to a CF100 flange (see Fig.~\ref{fig:cavity_thermics_drawing}). The aluminium holder is actively temperature regulated using two Peltier cells (Peltier (A) in \autoref{fig:cavity_thermics_drawing}) which dissipate heat through the CF100 flange. The temperature is monitored in-loop and out-of-loop via PT100 thermistors, with regulation provided by a temperature controller\footnote{Wavelength electronics TC5-lab.}, ensuring a \qty{0.1}{\kelvin}$ (\tau/\mathrm{s})^{-1/2}$ in-loop temperature stability. Passive thermal filtering is achieved by thermal isolation through the aluminium holder and Viton balls. An equivalent circuit model of the heat transfer from the vacuum chamber to the cavity, accounting for material properties and thermal radiation, showed a low-pass filter behavior with a \qty{5.4}{\micro\hertz} cut-off frequency. This agrees with the $\simeq \qty{8}{\hour}$ response time we observe experimentally. To further reduce the system sensitivity to the lab temperature, the vacuum chamber is temperature regulated using several Peltier cells (Peltiers (B) in \autoref{fig:cavity_thermics_drawing}). This secondary regulation reduced the out-of-loop residual temperature fluctuations by a factor~2. The residual out-of-loop temperature fluctuations, measured by their Allan deviation, stay below \qty{2e-3}{K} at all integration times.

\subsection{Coefficient of thermal expansion}
To characterize the cavity's frequency-temperature dependence,
we define a global coefficient of thermal expansion for the composite cavity as:

\begin{equation}
	\alpha_\mathrm{tot}(T) = \alpha_\mathrm{ZD}(T)\frac{L_\mathrm{ZD}}{L_\mathrm{tot}} + \alpha_\mathrm{PZT}(T)\frac{2 L_\mathrm{PZT}}{L_\mathrm{tot}} +
	\alpha_\mathrm{Kov}(T)\frac{2 L_\mathrm{Kov}}{L_\mathrm{tot}}.
\end{equation}

The temperature dependence of the coefficients of thermal expansion $\alpha_\mathrm{ZD}$ and $\alpha_\mathrm{Kov}$ is poorly documented. In particular, for Kovar, only a global value for the broad temperature range between \qty{25}{\celsius} and \qty{100}{\celsius} is given ($\alpha_\mathrm{Kov} = 5.86\times 10^{-6}~\mathrm{K^{-1}}$).
To estimate $\alpha_\mathrm{tot}(T)$, we thermally stabilized the system, then abruptly stepped the temperature set-point by \qtyrange{1}{5}{\kelvin}.
If $\alpha_\mathrm{tot}(T)$ is nearly constant over the explored temperature range, the frequency response of the system is primarily an exponential with a time constant of approximately eight hours, corresponding to the global response time. However, if $\alpha_\mathrm{tot}(T)$ changes sign over the temperature range, the frequency response exhibits non-monotonous behavior, with a turning point when the value of $\alpha_\mathrm{tot}(T)$ vanishes.

\begin{figure}[ht]
	\centering
	\includegraphics[width=\linewidth]{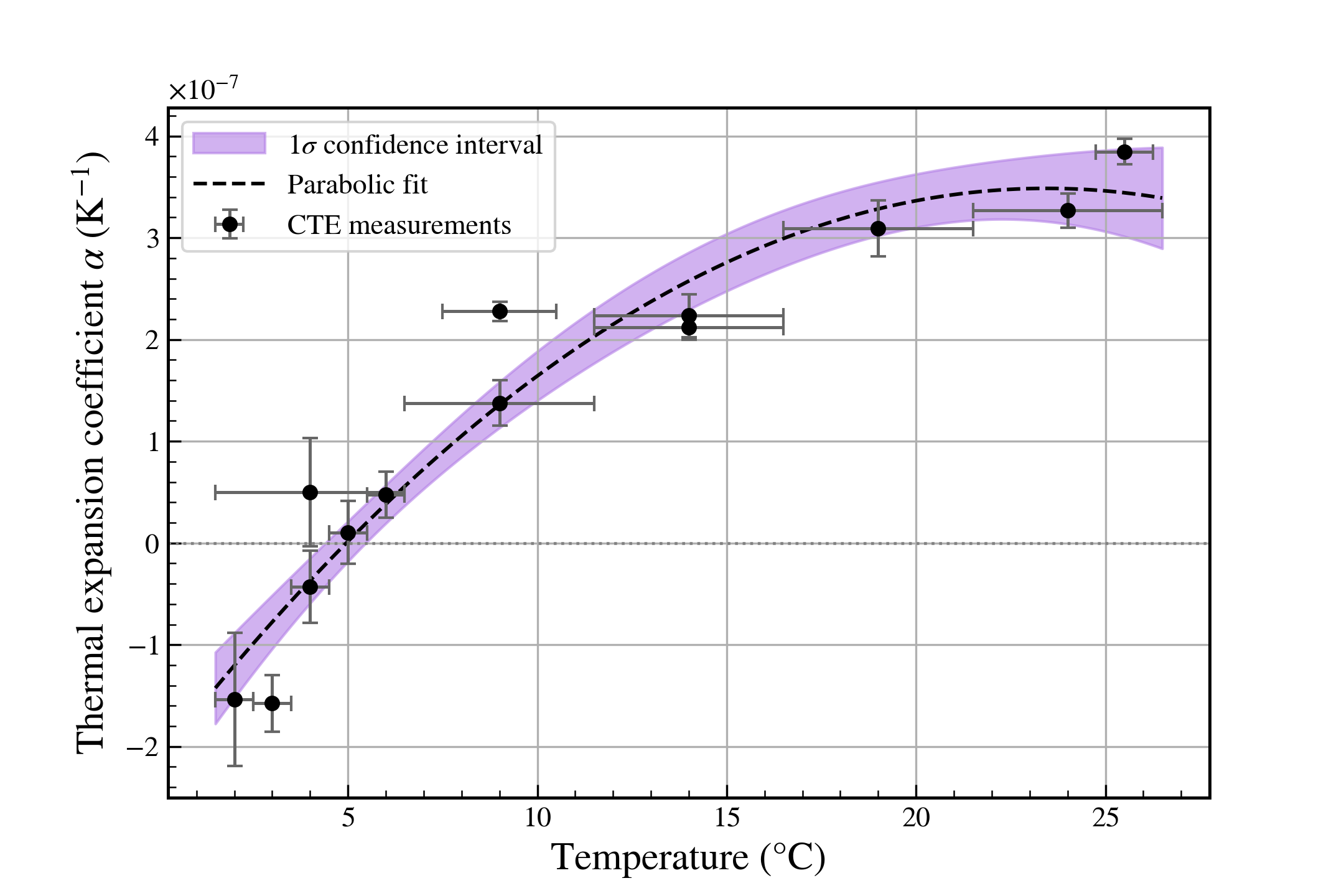}
	\caption{Cavity coefficient of thermal expansion measured by applying temperature steps to the cavity temperature set-point and measuring the subsequent changes in the \qty{578}{nm} laser frequency. A zero-crossing point is extracted at \qty{4.9\pm0.5}{\celsius}. Error bars are discussed in the main text.
	}
	\label{fig:cte_plot}
\end{figure}

Given potential temperature inhomogeneities during transient thermalization, we focused on the steady-state behavior. We measured the total frequency shift induced by temperature changes for various temperatures, as shown in  Fig.~\ref{fig:cte_plot}. The horizontal error bars correspond to the temperature range explored during the step while the vertical error bars are obtained by taking the standard deviation of the fit residuals when fitting exponential-decay-like curves to traces. The CTE clearly cancels near $T_0$, and a parabolic fit yielded a zero-crossing temperature  $T_0 = \qty{4.9\pm 0.5}{\celsius}$.  
Near $T_0$ the temperature dependence of $\alpha_\mathrm{tot}(T)$ can be approximated as $\alpha_\mathrm{tot}(T) = a_\mathrm{tot} (T-T_0)$ with $a_\mathrm{tot} = \qty{4.8 \pm 0.9e-8}{\kelvin^{-2}}$. For the following best measurements, we set the regulation temperature to $T_0$ to minimize thermal expansion effects.

\subsection{Stabilities}

\begin{figure}[ht]
	\centering
	\includegraphics[width=.8\linewidth]{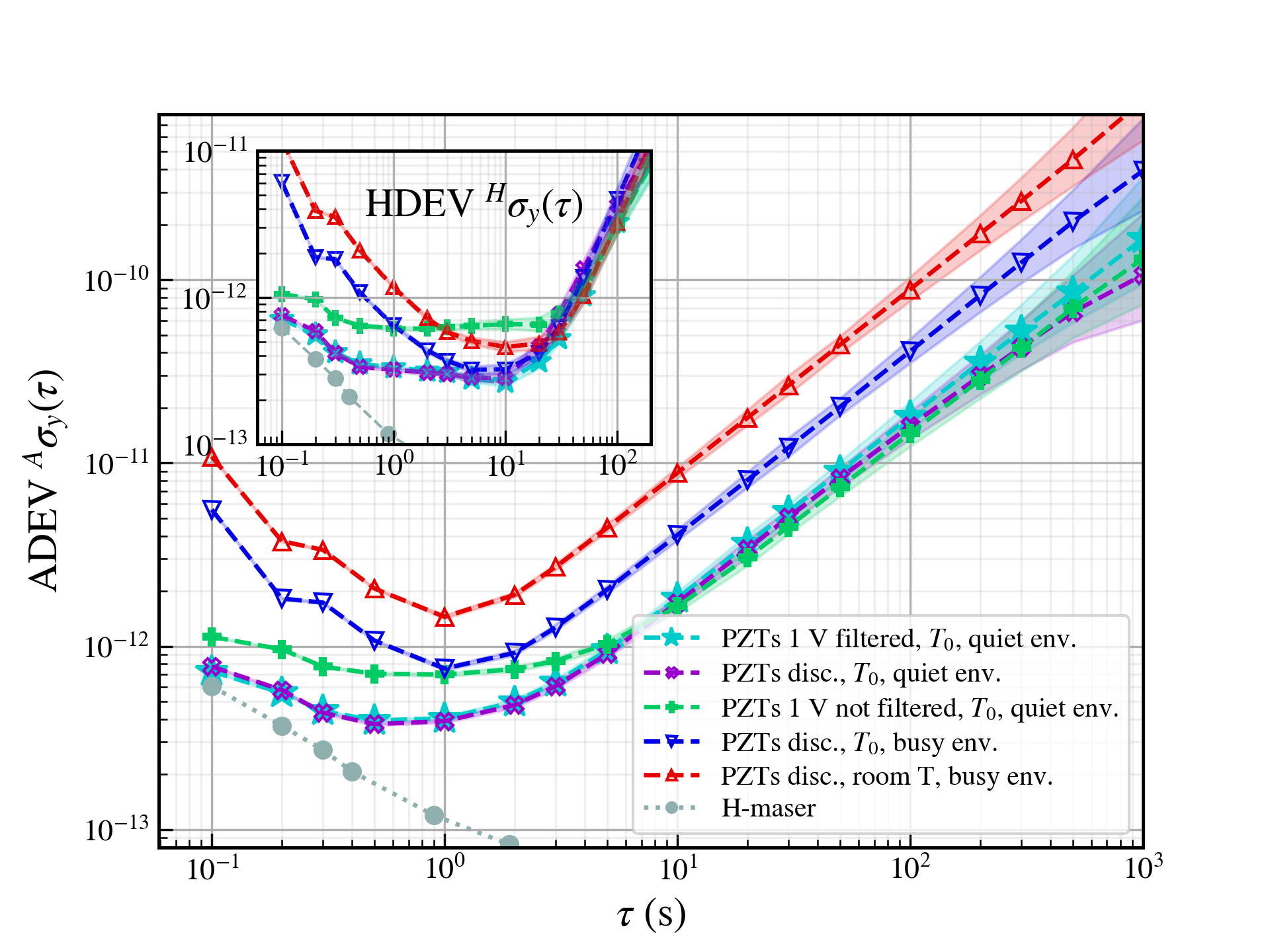}
	\caption{Fractional frequency stability of the tunable cavity measured against the H-maser, estimated using the unbiased Allan deviation under several conditions. Quiet environment (\textit{ie} atomic oven off): regulated at $T_0$ with PZTs disconnected (pink stars), regulated at $T_0$ with filtered PZTs (purple squares), and regulated at $T_0$ with non-filtered PZTs at \qty{1}{V} (green diamonds). Real operating conditions (atomic oven on): with PZTs disconnected at $T_0$ (blue lower triangles) and at room temperature (red upper triangles). Grey curve: fractional frequency stability of the H-maser measured against an ultra-stable cavity. Inset: unbiased Hadamard deviation of the same datasets, indicating the flicker frequency noise floor that could be achieved by suppressing frequency drift.}
	\label{fig:stab}
\end{figure}

Fig.~\ref{fig:stab} shows the fractional frequency stability measured against the H-maser via Allan deviation of the OFC beatnote. The stability of the H-maser measured against an ultra-stable cavity at the $10^{-15}$ level is indicated as the grey curve in Fig.~\ref{fig:stab}. We performed several measurement sets to isolate the PZT noise contribution. With PZTs disconnected or filtered at $T_0$, the cavity reaches \num{4e-13} at \qty{1}{s}. Without filtering, stability degrades by a factor $\sim 2$. Under real operating conditions (atomic oven on) at $T_0$, performance remains in the \num{e-13} range, while operating at room temperature with the oven on degrades it by a further factor of 2.

To isolate the effects of linear drifts from other noise sources, we also analyzed the frequency data with Hadamard deviation (HDEV) in the inset of Fig.~\ref{fig:stab}. The HDEV is by construction insensitive to linear drifts, and therefore mainly suppresses the effects of temperature fluctuations. For pure flicker frequency noise, this results in a $\tau^0$ slope in ADEV and HDEV, for which it can be shown that ${}^H\sigma_y (\tau) = \sqrt{\frac{8 \ln 2 - 3\ln 3}{6 \ln 2}}{}^A\sigma_y (\tau)=0.74 {}^A\sigma_y (\tau)$~\cite{Rubiola2022}. The HDEV plot emphasizes the degradation of the frequency stability induced by unfiltered power supplies, with a strong degradation of the flicker frequency noise.
The short-term measurements (for $\tau = \qty{100}{ms}$) are likely limited by the H-maser performance.

Overall, a fractional frequency stability of $\num{8e-13}$ at one second integration time is achievable for this  free-running cavity under voltage-controlled conditions, and a realistic experimental environment that corresponds to the operational regime of a superradiant laser. This level of stability is sufficiently low to avoid limiting the performance of a SR laser above $\num{8e-18}$ at one second - a level of stability that has not been achieved by any system to date, with possible improvements down to $\num{4e-18}$ if the influence of the atomic oven can be reduced. Other tunable cavities have been reported in the literature. For instance, \cite{mohle_highly_2013} reports longer tunable cavities with stabilities at the \num{e-15} level, but not designed for atomic physics. On the other hand, much shorter fiber Fabry-Perot cavities may give access to strong coupling in QED, at the price of a reduced stability \cite{saavedra_tunable_2021}. These results are summarized in Table~\ref{tab:comparison2}.

\begin{table}[H]
	\centering
	\resizebox{\textwidth}{!}{%
		\begin{tabular}{c|c|c|c}
			                                     & Our design                            & Möhle (PZN-PT) \cite{mohle_highly_2013} & Saavedra (triple-slot) \cite{saavedra_tunable_2021} \\
			\hline
			Length                               & \qty{56}{mm}                          & \qty{100}{mm}                           & \qty{93}{\micro m}                                  \\
			Finesse                              & 6920                                  & $\sim$300\,000                          & 99\,000                                             \\
			Freq. sensitivity (\unit{MHz/V})     & 233                                   & 2.2                                     & $6.1\times10^{3}$                                   \\
			CTE ($\times\qty{e-7}{\per\kelvin}$) & $\lesssim\pm0.1$ at \qty{5}{\celsius} & 3                                       & 5.5 (SiO$_2$, no zero-crossing)                     \\
			$\sigma_y$ (\qty{1}{s})              & \num{4e-13}                           & \num{8e-15}                             & $\sim \num{e-11}$                                   \\
		\end{tabular}}
	\caption{Comparison of our tunable FP cavity with designs from the literature.}
	\label{tab:comparison2}
\end{table}

\section{Conclusion}
\label{sec:conclusion}

We have realized a tunable optical FP cavity that is passively stable at the $\num{4e-13}$ level in typical operating conditions. This cavity features several optical access ports in order to couple atoms to its optical mode. We achieved this stability by operating at the CTE zero-crossing temperature of the composite cavity, $T_0$ at $\qty{4.9\pm 0.5}{\celsius}$.

A better suppression of temperature-induced drifts could further improve the frequency stability of the system. This could be implemented by applying feed-forward or PID correction to the PZT voltage using lab temperature measurements \cite{mangeot2016full, sabarianand_review_2020,janocha_real-time_2000}.

\section*{Data availability}

All the data used in this article, including designs, plot points, scripts, is available under a CC-BY-SA license at \cite{ruelle_temperature-insensitive_2026}.

\section*{Acknowledgements}
The authors would like to thank Pierre Roset and Jacques Millo for providing the ultra-stable optical frequency reference, as well as Yann Kersalé for careful reading of the manuscript.

\paragraph{Funding information}
This work has been supported by the EIPHI Graduate school (contract ANR-17-EURE-0002) and by the Bourgogne-Franche-Comté Region, by the ANR (project CONSULA, ANR-21-CE47-
0006-02), by First-TF Labex (contract ANR-10-LABX-48-01), by Oscillator-IMP Equipex
(contract ANR-11-EQPX-0033) and by the French Space Agency (CNES).

\bibliographystyle{SciPost_bibstyle}
\bibliography{SciPost_Example_BiBTeX_File.bib}

\end{document}